\documentstyle[12pt]{article}

\title{QUANTUM, GRAVITY, AND GEOMETRY}
\author{{\bf Ali Shojai}$^{1,2,}$\footnote{Email: SHOJAI@THEORY.IPM.AC.IR}
\\$^1$Department of Physics, Tarbiat Modares University,\\
P.O.Box 14155--4838, Tehran, IRAN,\\$^2$Institute for Studies in Theoretical 
Physics and Mathematics,\\P.O.Box 19395--5531, Tehran, IRAN.}
\date{}
\begin{document}
\maketitle
\newpage
\begin{abstract}
Recently\cite{GEO,STQ,CON}, it is shown that, the quantum effects of matter 
are well described by the conformal degree of freedom of the space--time metric.
On the other hand, it is a wellknown fact that according to Einstein's gravity 
theory, gravity and  geometry are interconnected. In the new quantum gravity 
theory\cite{GEO,STQ,CON}, matter quantum effects completely determine the conformal 
degree of freedom of the space--time metric, while the causal structure of the 
space--time is determined by the gravitational effects
of the matter, as well as the quantum effects through back reaction effects. 
This idea, previousely, is realized in the framework of scalar--tensor theories. 
In this work, it is shown that quantum gravity theory can also be realized as a 
purely metric theory. Such a theory is developed, its consequences and 
its properties are investigated. The theory is applied, then, to black holes and 
the radiation--dominated universe. It is shown that the initial singularity 
can be avoided.
\end{abstract}
\section{INTRODUCTION AND SURVEY}
This paper concerns with the generalization of some specific approach to quantum 
gravity presented in \cite{GEO,STQ}. We shall construct a purely tensor theory 
instead of the scalar--tensor theory presented in \cite{STQ}. Our approach would 
be applicable to purely quantum gravity effects as well as to quantum effects of 
matter in a curved space--time. First of all we shall breifly review the motivation 
and basic points of the quantum gravity theory presented in \cite{GEO,STQ}.

In this century, physicists have been departed from 19th century physics, in 
two ways. The first was the generalization and bringing the old idea of 
{\it frame independence\/} or 
{\it general covariance\/}, in a manifest form. The result of this effort was the 
pioneer general relativity theory, in which the gravitational effects of matter 
are identified with the geometry of the space--time.  The enigmatic character of 
this theory is just the above--mentioned property, i.e., the interconnection of 
gravity and general covariance. When one tries to make a general
covariant theory, one is forced to include gravity. The main root of this 
interconnection is the {\it equivalance principle}. According to the equivalance 
principle, it is possible to go to a frame in which gravity is locally absent, 
and thus special theory of relativity is applicable locally. Now using the general 
covariance and writting down anything in a general frame, we will get the general 
relativity theory.\cite{WEI}

The second was the investigation of the quantal behaviour of matter, that  leads 
to the {\it quantum theory}. According to which a great revolution is appeared 
in physics. In order to explain the atomic world, quantum theory threw out 
two essensial classical concepts, the {\it principle of causality\/} and 
{\it the dogma of formulation of  physics in terms of motion in the 
space--time} (motion dogma). The first one is violated during a measurement 
process, while the second
does not exist at any time. 

After appearance of quantum mechanics, it was proved that not only the 
ordinary particles show quantal behaviour but also mediators of the fundamental 
forces do so. In this way quantum electrodynamics, quantum chromodynamics and 
quantum flavourdynamics were born. But the construction of quantum gravitodynamics 
or quantum gravity, leads to several structural and conceptual 
difficulties\cite{DIF}. For example, the meaning of the wavefunction of the universe 
is a conceptual difficulty while the time independence of the wavefunction 
is a structural one.

In contrast to general theory of relativity which is the best theory for gravity, 
standard quantum mechanics is not the only satisfactory way of understanding the 
quantal behaviour of matter.
One of the best theories explaining the quantal behaviour of matter but remaining 
faithful to the principle of causality and the motion dogma, is the de-Broglie--Bohm 
quantum theory.\cite{BOH}
According to this theory, all the enigmatic quantal behaviour of the matter is 
resulted from a self--interaction of the particle. In fact, any particle 
exerts a {\it quantum force\/}
on itself which can be expressed in terms of a {\it quantum potential\/}
and which is derived from the particle wavefunction.

In the non--relativistic de-Broglie--Bohm theory, the quantum potential is 
given by\cite{BOH}:
\begin{equation}
{\cal Q}=-\frac{\hbar^2}{2m}\frac{\nabla^2|\Psi|}{|\Psi|}
\end{equation}
in which $\Psi$ is the particle's wavefunction satisfying an appropriate
wave equation. (in this case the nonrelativistic Schr\"odinger equation) 
The particle's trajectory
then can be derived from Newton's law of motion in which 
in addition to the classical force $-\vec{\nabla}V$, the quantum force
$-\vec{\nabla}{\cal Q}$ is also present. In this way the de-Broglie--Bohm 
quantum theory presents a formulation of physics in terms of motion in the
space--time.

The celebrated property of the de-Broglie--Bohm quantum theory is the 
following property. {\it At anytime, even when a measurement is done, the 
particle is on the trajectory given by Newton's law of motion, including 
the quantum force}. During a measurement, the system is in fact a many--body
system (including the particle itself, the probe particle, and the 
registrating system particles). When one writes down the appropriate equation
of motion of all the particles\footnote{It must be noted that the quantum 
potential for a many--body system is given by:
\[ {\cal Q}=-\frac{\hbar^2}{2}\sum_{i=1}^N\frac{1}{m_i}
\frac{\nabla_i^2|\Psi|}{|\Psi|}\]
where $m_i$ is the mass of $i$th particle and $\vec{\nabla}_i$ represents
differentiating with respect to the position of the $i$th particle, and $N$
is the number of particles.}, and when one considers the very fact that we
know nothing about the initial conditions of the registerating system
particles, one sees how the projection postulate of quantum mechanics
came out\cite{BOH}. According to which the result of any measurement is one of the 
eigenvalues of the operator related to the measured quantity with some
calculable probability distribution.\cite{BOH}

Another important property of the de-Broglie--Bohm quantum theory is that using
the Schr\"odinger equation and the Newton's law of motion, one observes that
the phase of the wavefunction is proportional to the Hamilton--Jacobi
function of the system. In fact if one set $\Psi=\sqrt{\rho}\exp[iS/\hbar]$
then one arrives at:
\begin{equation}
\frac{\partial S}{\partial t}+\frac{|\vec{\nabla}S|^2}{2m}+V+{\cal Q}=0
\label{nhj}
\end{equation}
\begin{equation}
\frac{\partial \rho}{\partial t}+\vec{\nabla}\cdot\left (\rho
\frac{\vec{\nabla}S}{m}\right )=0
\end{equation}
The first of these equations (the Hamilton--Jacobi equation) is identical to the 
Newton's second law. It is in fact the energy condition:
\begin{equation}
E=\frac{|\vec{p}|^2}{2m}+V+{\cal Q}
\end{equation}
Remember that in the Hamilton--Jacobi theory\cite{GOL} $-\partial S/\partial t=E$ and
$\vec{\nabla}S=\vec{p}$. The second equation is then the continuity equation 
of an hypothetical ensemble of the particle.

The relativistic extension of the Bohmian quantum theory 
is straightforward\cite{BOH}.
All we must to do is to generalize the relativistic energy equation:
\begin{equation}
\eta_{\mu\nu}P^\mu P^\nu =m^2c^2
\end{equation}
to :
\begin{equation}
\eta_{\mu\nu}P^\mu P^\nu =m^2c^2+{\cal Q}={\cal M}^2c^2
\end{equation}
where:
\begin{equation}
{\cal Q}=\hbar^2\frac{\Box|\Psi|}{|\Psi|}
\end{equation}
or
\begin{equation}
{\cal M}^2=m^2\left ( 1+\alpha\frac{\Box|\Psi|}{|\Psi|}\right );
\ \ \ \ \ \ where\ \ \ \alpha=\frac{\hbar^2}{m^2c^2}
\label{x1}
\end{equation}
In fact one can derive this by setting $\Psi=\sqrt{\rho}e^{iS/\hbar}$ in the 
Klein--Gordon equation, and separating the real and imaginiary parts. The
result is the relativistic Hamilton--Jacobi equation:
\begin{equation}
\label{rhj}
\eta_{\mu\nu}\partial^\mu S\partial^\nu S={\cal M}^2c^2
\end{equation}
which is identical to (\ref{nhj}) (Note that $P^\mu=-\partial^\mu S$ ) 
and the continuity equation
\begin{equation}
\partial_\mu\left (\rho\partial^\mu S\right )=0
\end{equation}
An important problem about the relativistic de-Broglie--Bohm theory is that 
the mass square defined by equation (\ref{x1}), is not positive--definite\cite{BOH}.
Later, we shall show that this problem would be solved in our present theory.

The generilization of (\ref{rhj}) to an arbitrary curved space--time with
metric $g_{\mu\nu}$ is:
\begin{equation}
\label{chj}
g_{\mu\nu}P^\mu P^\nu={\cal M}^2c^2
\end{equation}
The particle trajectory can be derived from the guidance relation:
\begin{equation}
P^\mu={\cal M}\frac{dx^\mu}{d\tau}
\end{equation}
or by differentiating (\ref{chj}) leading to the Newton's equation of motion:
\begin{equation}
\label{qg}
{\cal M}\frac{d^2x^\mu}{d\tau^2}+{\cal M}\Gamma^\mu_{\nu\kappa}u^\nu u^\kappa
=(g^{\mu\nu}-u^\mu u^\nu)\nabla_\nu{\cal M}
\end{equation}
The essential observation in references \cite{GEO,STQ,CON} is that equation
(\ref{chj}) can be transformed to the classical (i.e. non quantum) one:
\begin{equation}
g_{\mu\nu}P^\mu P^\nu=m^2c^2
\end{equation}
via a conformal transformation:
\begin{equation}
\label{cf}
g_{\mu\nu}\longrightarrow \widetilde{g}_{\mu\nu}=\left ( \frac{{\cal M}^2}{m^2}\right )^{-1}
g_{\mu\nu}
\end{equation}
In an equivalant manner, equation (\ref{qg}) reduces to the standard geodesic equation
via the above conformal transformation.

The physical conclusion from this observation is the following:
{\it Suppose in the metric $g_{\mu\nu}$ there is no quantal effect at all (i.e. the 
trajectory of the particle is a geodesic), the quantal effect can be brought
in via a specific conformal transformation as above.}

In ref.\cite{GEO} the authors as a first step towards the formulation of 
the above conclusion, introduced the quantum conformal degree of
freedom via the method of lagrange multipliers. In this way they have a set of
equations of motion describing the background metric, the conformal degree
of freedom and the particle trajectory. As a brief introduction to their work
we present here their action principle and equations of motion. The action is\cite{GEO}:
\[ {\cal A}[\overline{g}_{\mu \nu},\Omega,S,\rho,\lambda]
=\frac{1}{2\kappa}\int d^4x\sqrt{-\overline{g}}\left 
(\overline{{\cal R}}\Omega^2-6\overline{\nabla}_{\mu}
\Omega\overline{\nabla}^{\mu}\Omega\right ) \]
\[ +\int d^4x \sqrt{-\overline{g}}\left ( \frac{\rho}{m}\Omega^2
\overline{\nabla}_{\mu}S\overline{\nabla}^{\mu}S -m\rho\Omega^4\right)\]
\begin{equation}
+\int d^4x \sqrt{-\overline{g}}\lambda \left [ \Omega^2-\left 
( 1+\frac{\hbar^2}{m^2}\frac{\stackrel{-}{\Box}\sqrt{\rho}}{\sqrt{\rho}}
\right ) \right ]
\end{equation}
in which any overlined quantity is calculated in the background metric, 
$\lambda$ is the lagrange multiplier, and $\Omega^2$ is the conformal factor. 
The equations of motion are\cite{GEO}:
\begin{enumerate}
\item The equation of motion for $\Omega$:
\begin{equation}
\overline{{\cal R}}\Omega+6\stackrel{-}{\Box}\Omega+
\frac{2\kappa}{m}\rho \Omega \left ( \overline{\nabla}_{\mu}S
\overline{\nabla}^{\mu}S-2m^2\Omega^2\right )+2\kappa \lambda\Omega=0
\end{equation}
\item The continuity equation for particles:
\begin{equation}
\overline{\nabla}_{\mu}\left (\rho \Omega^2 \overline{\nabla}^{\mu}S \right )=0
\end{equation}
\item The equation of motion for particles:
\begin{equation}
\left ( \overline{\nabla}_{\mu}S \overline{\nabla}^{\mu}S 
-m^2\Omega^2\right )\Omega^2\sqrt{\rho}+\frac{\hbar^2}{2m}\left 
[ \stackrel{-}{\Box}\left (\frac{\lambda}{\sqrt{\rho}}\right ) -
\lambda\frac{\stackrel{-}{\Box}\sqrt{\rho}}{\rho}\right ]=0
\end{equation}
\item The modified Einstein equations for $\overline{g}_{\mu \nu}$:
\[ \Omega^2\left [ \overline{{\cal R}}_{\mu \nu}-\frac{1}{2}
\overline{g}_{\mu \nu}\overline{{\cal R}}\right ]
-\left [ \overline{g}_{\mu \nu}\stackrel{-}{\Box} -
\overline{\nabla}_{\mu}\overline{\nabla}_{\nu}\right ]
\Omega^2 -6 \overline{\nabla}_{\mu}\Omega \overline{\nabla}_{\nu}
\Omega+3\overline{g}_{\mu \nu}\overline{\nabla}_{\alpha}\Omega 
\overline{\nabla}^{\alpha}\Omega\]
\[ +\frac{2\kappa}{m}\rho\Omega^2 \overline{\nabla}_{\mu}S 
\overline{\nabla}_{\nu}S-\frac{\kappa}{m}\rho \Omega^2\overline{g}_{\mu \nu} 
\overline{\nabla}_{\alpha}S \overline{\nabla}^{\alpha}S +\kappa m 
\rho \Omega^4 \overline{g}_{\mu \nu}\]
\begin{equation}
+\frac{\kappa\hbar^2}{m^2}\left [ \overline{\nabla}_{\mu}\sqrt{\rho}
\overline{\nabla}_{\nu}\left ( \frac{\lambda}{\sqrt{\rho}}\right )
 \overline{\nabla}_{\nu}\sqrt{\rho}\overline{\nabla}_{\mu}\left (
 \frac{\lambda}{\sqrt{\rho}}\right )\right]
 -\frac{\kappa\hbar^2}{m^2}\overline{g}_{\mu \nu}
\overline{\nabla}_{\alpha}\left [\lambda 
\frac{\overline{\nabla}^{\alpha}\sqrt{\rho}}{\sqrt{\rho}}\right ]=0
\end{equation}
\item The constraint equation:
\begin{equation}
\Omega^2=1+\frac{\hbar^2}{m^2}\frac{\stackrel{-}{\Box}\sqrt{\rho}}{\sqrt{\rho}}
\end{equation}
\end{enumerate}
If one assumes that anything can be expanded in terms of powers 
of Planck's constant, one gets zero for $\lambda$ and the equations of motion are\cite{GEO}:
\begin{equation}
\overline{\nabla}_{\mu}\left (\rho \Omega^2 \overline{\nabla}^{\mu}S \right )=0
\end{equation}
\begin{equation}
\overline{\nabla}_{\mu}S \overline{\nabla}^{\mu}S =m^2\Omega^2
\end{equation}
\[ \Omega^2\left [ \overline{{\cal R}}_{\mu \nu}-
\frac{1}{2}\overline{g}_{\mu \nu}\overline{{\cal R}}\right ]
-\left [ \overline{g}_{\mu \nu}\stackrel{-}{\Box} -\overline{\nabla}_{\mu}
\overline{\nabla}_{\nu}\right ]
\Omega^2 -6 \overline{\nabla}_{\mu}\Omega \overline{\nabla}_{\nu}\Omega
+3\overline{g}_{\mu \nu}\overline{\nabla}_{\alpha}\Omega 
\overline{\nabla}^{\alpha}\Omega\]
\begin{equation}
+\frac{2\kappa}{m}\rho\Omega^2 \overline{\nabla}_{\mu}S
\overline{\nabla}_{\nu}S-\frac{\kappa}{m}\rho \Omega^2\overline{g}_{\mu
\nu} \overline{\nabla}_{\alpha}S \overline{\nabla}^{\alpha}S 
+\kappa m \rho \Omega^4 \overline{g}_{\mu \nu}=0
\end{equation}
\begin{equation}
\Omega^2=1+\alpha\frac{\stackrel{-}{\Box}\sqrt{\rho}}{\sqrt{\rho}}
\end{equation}
In ref.\cite{GEO} some features of the above formulation is presented. For
example the solution for a Rabertson--Walker metric is derived, and the 
formulation is extended to contain the radiation quantum effects. Also nonlocal
effects in the framework of such theory is investigated\cite{NQG}. We shall 
come back to this point later.

As a next step, currently\cite{STQ,CON} it is shown that there is a close connection
with scalar--tensor theories. In fact it is possible to set up a scalar--tensor
theory in which there is no need to use the method of lagrange multipliers
and in which the correct conformal degree of freedom, i.e. one given by
quantum effects as in equation (\ref{cf}), is derived using the equations of motion.
The action functional for this theory is\cite{STQ}:
\begin{equation}
{\cal A}=\int d^4x \left \{ \phi {\cal R}-\frac{\omega}{\phi}\nabla^\mu\phi
\nabla_\mu\phi+2\Lambda\phi+{\cal L}_m\right \}
\end{equation}
\begin{equation}
{\cal L}_{m}=\frac{\rho}{m}\nabla_\mu S\nabla^\mu S-\rho m\phi-\Lambda(1+{\cal Q})^2
\end{equation}
in which $\Lambda$ is the cosmological constant.
The equations of motion are\cite{STQ}:
\begin{equation}
\phi=1+Q-\frac{\alpha}{2}\Box Q
\label{YYY}
\end{equation}
\begin{equation}
\nabla^\mu S\nabla_\mu S=m^2\phi-\frac{2\Lambda m}{\rho}(1+Q)(Q-\widetilde{Q})
+\frac{\alpha\Lambda m}{\rho}\left ( \Box Q -2\nabla_\mu Q\frac{\nabla^\mu 
\sqrt{\rho}}{\sqrt{\rho}}\right )
\label{ZZZ}
\end{equation}
\begin{equation}
\nabla^\mu(\rho\nabla_\mu S)=0
\end{equation}
\begin{equation}
{\cal G}^{\mu\nu}-\Lambda g^{\mu\nu}=-\frac{1}{\phi}{\cal T}^{\mu\nu}
-\frac{1}{\phi}[\nabla^\mu\nabla^\nu-g^{\mu\nu}\Box ]\phi+\frac{\omega}{\phi^2}
\nabla^\mu\phi\nabla^\nu\phi-\frac{1}{2}\frac{\omega}{\phi^2}g^{\mu\nu}
\nabla^\alpha\phi\nabla_\alpha\phi
\label{TTT}
\end{equation}
where
\begin{equation}
\widetilde{{\cal Q}}=\alpha\frac{\nabla_\mu\sqrt{\rho}\nabla^\mu\sqrt{\rho}}{\rho}
\end{equation}

Both the above mentioned theories\cite{GEO,STQ}, have a problem. 
In these theories, it is assumed that one deals with an ensemble of similar 
particles with density $\rho$. In Bohm's theory, the quantum potential 
exists for a single particle as well as for an ensemble. In the case 
of a single particle, the interpretation of the quantum potential is in terms of 
an hypothetical ensemble. 
Note that in the above theories, the ensemble is a real one, not an hypothetical 
one, because, the energy--momentum tensor of the ensemble is appeared and 
has physical effects. As we shall show later, in this paper we have 
solved this problem and the theory would work both for a single particle and 
for an ensemble.

In addition to the above problem there is still another one. The  theory
considers only the quantal effects of matter, the pure quantum gravity effects 
are not included.

In this paper we shall show that it is possible to make a pure tensor theory for quantum 
gravity. As a result we shall show that the correct quantum conformal degree 
of freedom would be achieved, and that the theory works for a particle as well as
for a real ensemble of the particle under consideration and that it includes the
pure quantum gravity effects. We shall do all of these by trying to write the quantum
potential terms in term of geometrical parameters not in terms of ensemble 
properties.
\section{A TOY QUANTUM--GRAVITY THEORY}
In this section we first examine how we can translate the quantum potential in a complete
geometrical manner, i.e. we write it in a form that there is no explicite reference to matter
parameters. Only after using the field equations one deduce the original form of the
quantum potential. This has the advantage that lets our theory work both for a single
particle and an ensemble. Next we write a special field equation as a toy theory and extract
some of its consequences.
\subsection{Geometry of the Quantum Conformal Factor}
In order to construct a purely metric theory for quantum gravity, 
let us first examine the geometrical properties of the conformal factor given by:
\begin{equation}
g_{\mu\nu}=e^{4\Lambda}\eta_{\mu\nu};\ \ \ \ \ \ \ \ e^{4\Lambda}=\frac{{\cal M}^2}{m^2}
=1+\alpha\frac{\Box_\eta\sqrt{\rho}}{\sqrt{\rho}}
\label{bala}
\end{equation}
where $\eta_{\mu\nu}$ is the flat space--time metric, $\Box_\eta$ is the dalambertian 
operator evaluated using the $\eta_{\mu\nu}$ metric.

If one evaluates the Einstein's tensor for the above metric, one gets:
\begin{equation}
{\cal G}_{\mu\nu}=4g_{\mu\nu}e^\Lambda\Box_\eta e^{-\Lambda}+2e^{-2\Lambda}
\partial_\mu\partial_\nu e^{2\Lambda}
\end{equation}
Now replacing $|{\cal T}|$ for $\rho$ in equation (\ref{bala}) 
(as it is true for a dust)\footnote{The
absolute value sign is introduced to make the square root always meaningful.}:
\begin{equation}
\frac{\Lambda}{\alpha}=e^{-4\Lambda}\frac{\Box\sqrt{|{\cal T}|}-
4(\nabla\Lambda)^2}{\sqrt{|{\cal T}|}}
\end{equation}
and expanding anything up to first order in $\alpha$ one gets:
\begin{equation}
\Lambda=\alpha\frac{\Box\sqrt{|\cal T|}}{\sqrt{|\cal T|}}
\end{equation}
As an ansatz, we suppose that in the presense of gravitational effects, 
the field equations are:
\begin{equation}
{\cal R}_{\mu\nu}-\frac{1}{2}{\cal R}g_{\mu\nu} = \kappa {\cal T}_{\mu\nu} + 4 g_{\mu\nu} e^\Lambda \Box e^{-\Lambda} + 2 e^{-2\Lambda} \nabla_\mu \nabla_\nu e^{2\Lambda}
\label{x2}
\end{equation}
This equation is written in such a way that in the limit ${\cal T}_{\mu\nu}\rightarrow 0$ 
the solution (\ref{bala}) achieved.

Making the trace of the above equation one gets:
\begin{equation}
-{\cal R}=\kappa {\cal T}-12\Box\Lambda+24(\nabla\Lambda)^2
\end{equation}
which has the iterative solution:
\begin{equation}
\kappa{\cal T}=-{\cal R}+12\alpha\Box\left ( \frac{\Box 
\sqrt{{\cal R}}}{\sqrt{{\cal R}}}\right ) +\cdots
\end{equation}
leading to:
\begin{equation}
\Lambda=\alpha \frac{\Box \sqrt{|{\cal T}|}}{\sqrt{|{\cal T}|}}\simeq \alpha
\frac{\Box \sqrt{|{\cal R}|}}{\sqrt{|{\cal R}|}}
\end{equation}
\subsection{Field Equations of a Toy Quantum Gravity}
In the previous subsection we have learned that ${\cal T}$ can be replaced with
${\cal R}$ in the expression for the quantum potential or for the conformal factor
of the space--time metric. This replacement is in fact an important improvement,
because the explicit reference to ensemble density is removed. This lets the theory
to work for both a single particle and an ensemble.

So with a glance at equation (\ref{x2}) for our toy quantum--gravity theory, 
we assume the following field equations:
\begin{equation}
{\cal G}_{\mu\nu}-\kappa{\cal T}_{\mu\nu}-{\cal Z}_{\mu\nu\alpha\beta}
\exp\left [\frac{\alpha}{2}\Phi\right ]
\nabla^\alpha\nabla^\beta
\exp\left [-\frac{\alpha}{2}\Phi\right ]
=0
\label{xz}
\end{equation}
where
\begin{equation}
{\cal Z}_{\mu\nu\alpha\beta}=2\left [ g_{\mu\nu}g_{\alpha\beta}-g_{\mu\alpha}
g_{\nu\beta}\right ]
\end{equation}
\begin{equation}
\Phi=\frac{\Box\sqrt{|{\cal R}|}}{\sqrt{|{\cal R}|}}
\end{equation}
Note that the number $2$ and the minus sign of the second term of the last 
equation are chosen so that the energy equation derived later be correct.
It would be very useful to take the trace of equation (\ref{xz}):
\begin{equation}
{\cal R}+\kappa{\cal T}+6\exp[\alpha\Phi/2]\Box\exp[-\alpha\Phi/2]=0
\label{x4}
\end{equation}
In fact this equation represents the connection of Ricci scalar curvature
of the space--time and the trace of matter energy--momentum tensor. This
is the analogeous of the Einstein's relation ${\cal R}=-\kappa{\cal T}$.
In the cases when a pertuerbative solution is admitted, i.e. when we {\it can\/}
expand anything in terms of powers of $\alpha$, one can find the relation between
${\cal R}$ and ${\cal T}$ perturbatively. In the zeroth approaximation
one has the classical relation:
\begin{equation}
{\cal R}^{(0)}=-\kappa{\cal T}
\end{equation}
As a better approaximation up to first order in $\alpha$, one gets:
\begin{equation}
{\cal R}^{(1)}=-\kappa{\cal T}
-6\exp[\alpha\Phi^{(0)}/2]\Box\exp[-\alpha\Phi^{(0)}/2]
\end{equation}
where:
\begin{equation}
\Phi^{(0)}=\frac{\Box\sqrt{|{\cal T}|}}{\sqrt{|{\cal T}|}}
\end{equation}
A more better result can be obtained in the second order as:
\begin{equation}
{\cal R}^{(2)}=-\kappa{\cal T}
-6\exp[\alpha\Phi^{(0)}/2]\Box\exp[-\alpha\Phi^{(0)}/2]
-6\exp[\alpha\Phi^{(1)}/2]\Box\exp[-\alpha\Phi^{(1)}/2]
\end{equation}
with:
\begin{equation}
\Phi^{(1)}=\frac{\Box\sqrt{|-\kappa{\cal T}-6
\exp[\alpha\Phi^{(0)}/2]\Box\exp[-\alpha\Phi^{(0)}/2]
|}}{\sqrt{|-\kappa{\cal T}
\exp[\alpha\Phi^{(0)}/2]\Box\exp[-\alpha\Phi^{(0)}/2]
|}}
\end{equation}

The energy relation can be obtained via taking the four divergance of 
the field equations. Using the fact that the divergance of the 
Einstein's tensor is zero, one gets:
\begin{equation}
\kappa\nabla^\nu{\cal T}_{\mu\nu}=\alpha{\cal R}_{\mu\nu}\nabla^\nu\Phi-
\frac{\alpha^2}{4} \nabla_\mu(\nabla\Phi)^2+\frac{\alpha^2}{2}\nabla_\mu\Phi\Box\Phi
\end{equation}
For a dust with:
\begin{equation}
{\cal T}_{\mu\nu}=\rho u_\mu u_\nu
\end{equation}
and assuming the conservation law for mass:
\begin{equation}
\nabla^\nu\left ( \rho{\cal M}u_\nu\right )=0
\end{equation}
up to first order in $\alpha$ one arrives at:
\begin{equation}
\frac{\nabla_\mu{\cal M}}{{\cal M}}=-\frac{\alpha}{2}\nabla_\mu\Phi
\end{equation}
or:
\begin{equation}
{\cal M}^2=m^2\exp (-\alpha\Phi)
\label{x3}
\end{equation}
where $m$ is some integration constant. Note that the first two terms of the above equation is 
in accordance with the relation (\ref{x1}). Higher order terms which are smaller than the first
two terms, are intereseting, because ${\cal M}^2$ given by (\ref{x3}) is positive--definite, while
${\cal M}^2$ given by (\ref{x1}) is not. So in this way an important problem of the de-Broglie--Bohm
theory is solved.
\section{SOME GENERAL SOLUTIONS}
In this section we shall study some general solutions of the field equation
(\ref{xz}).
\subsection{Conformally Flat Solution}
Suppose we search for a solution which is conformally flat, and that the 
conformal factor is near unity. Such a solution is of the form:
\begin{equation}
g_{\mu\nu}=e^{2\Lambda}\eta_{\mu\nu};\ \ \ \ \ \  \Lambda\ll 1
\end{equation}
As a result one can derive the following relations:
\begin{equation}
g=-e^{8\Lambda}=-1-8\Lambda
\end{equation}
\begin{equation}
{\cal R}_{\mu\nu}=\eta_{\mu\nu}\Box\Lambda+2\partial_\mu\partial_nu\Lambda
\end{equation}
\begin{equation}
{\cal R}=6\Box\Lambda
\end{equation}
\begin{equation}
{\cal G}_{\mu\nu}=2\partial_\mu\partial_\nu\Lambda -2\eta_{\mu\nu}\Box\Lambda
\end{equation}
In order to solve for $\Lambda$ one can use the relation (\ref{x4}), and solve it
iteratively as it is disscused in the previous section. The result is:
\begin{equation}
{\cal R}^{(0)}=-\kappa {\cal T}=6\Box\Lambda^{(0)}\Longrightarrow 
\Lambda^{(0)}=-\frac{\kappa}{6}\Box^{-1}{\cal T}
\end{equation}
\[
{\cal R}^{(1)}=-\kappa {\cal T}+3\alpha\Box\frac{\Box\sqrt{|{\cal T}|}}{\sqrt{|{\cal T}|}}
=6\Box\Lambda^{(1)}\Longrightarrow 
\]
\begin{equation}
\Lambda^{(1)}=-\frac{\kappa}{6}\Box^{-1}{\cal T}+\frac{\alpha}{2}
\frac{\Box\sqrt{|{\cal T}|}}{\sqrt{|{\cal T}|}}
\end{equation}
and so on. Thus:
\begin{equation}
\Lambda=\underbrace{-\frac{\kappa}{6}\Box^{-1}{\cal T}}_{\rm pure\ gravity}
\underbrace{+\frac{\alpha}{2}
\frac{\Box\sqrt{|{\cal T}|}}{\sqrt{|{\cal T}|}}}_{\rm pure\ quantum}
+{\rm higher\ terms\ including\ gravity-quantum\ interactions}.
\end{equation}
where $\Box^{-1}$ represents the inverse of the dalambertian operator. Note that the solution 
is in complete agreement with de-Broglie--Bohm theory.
\subsection{Conformally Quantic Solution}
As a generalization of the solution found in the previous subsection, suppose we set:
\begin{equation}
g_{\mu\nu}=e^{2\Lambda}\overline{g}_{\mu\nu}=(1+2\Lambda)\overline{g}_{\mu\nu};
\ \ \ \ \Lambda\ll 1
\end{equation}
One can evaluates the following relations:
\begin{equation}
{\cal R}_{\nu\rho}=\overline{{\cal R}}_{\nu\rho}+\overline{g}_{\mu\nu}
\stackrel{-}{\Box}\Lambda
+2\left ( \overline{\nabla}_\nu\overline{\nabla}_\rho\Lambda+\overline{g}_{\nu\rho}
\overline{\nabla}_\alpha\Lambda
\overline{\nabla}^\alpha\Lambda-\overline{\nabla}_\nu\Lambda\overline{\nabla}_\rho
\Lambda\right )
\end{equation}
\begin{equation}
{\cal R}=e^{-2\Lambda}\left ( \overline{{\cal R}}+6\stackrel{-}{\Box}\Lambda+6
\overline{\nabla}_\alpha
\Lambda\overline{\nabla}^\alpha\Lambda\right )
\end{equation}
\begin{equation}
{\cal G}_{\nu\rho}=\overline{{\cal G}}_{\nu\rho}-2\overline{g}_{\nu\rho}
\stackrel{-}{\Box}\Lambda
+2\overline{\nabla}_\nu\overline{\nabla}_\rho\Lambda
\end{equation}
On using these relations in the field equations (\ref{xz}) one gets the following solution:
\begin{equation}
\overline{{\cal G}}_{\mu\nu}=\kappa\overline{\cal T}_{\mu\nu};\ \ \ \ \Lambda=
\frac{\alpha}{2}\Phi
\end{equation}
provided the energy--momentum tensor be conformally invariant. So under 
this condition we have:
\begin{equation}
g_{\mu\nu}^{\rm quantum+gravity}=(1+\alpha\Phi)g_{\mu\nu}^{\rm gravity}
\end{equation}
\subsection{Conformally Highly Quantic Solution}
Now we can generalize the result of the previous subsection. Suppose 
in the overlined metric there is
no quantum effect, so that:
\begin{equation}
\overline{{\cal G}}_{\mu\nu}=\kappa\overline{{\cal T}}_{\mu\nu}
\end{equation}
and assuming that the quantum effects could bring in via a conformal transformation like:
\begin{equation}
g_{\mu\nu}=e^{2\Lambda}\overline{g}_{\mu\nu}
\end{equation}
Using the field equations (\ref{xz}) and the transformation properties 
of the Einstein's equation one gets:
\begin{equation}
2\stackrel{-}{\Box}\Lambda+2\overline{\nabla}_\alpha \Lambda
\overline{\nabla}^\alpha \Lambda =\alpha\stackrel{-}{\Box}\Phi+
2\alpha\overline{\nabla}_\alpha \Phi\overline{\nabla}^\alpha 
\Lambda -\frac{\alpha^2}{2}\overline{\nabla}_\alpha \Phi \overline{\nabla}^\alpha \Phi
\end{equation}
which has the solution:
\begin{equation}
\Lambda=\frac{\alpha}{2}\Phi
\end{equation}
In the above solution it is assumed that the energy--momentum tensor is 
either zero or conformally invariant.
So, under this condition, no matter how large is the quantum effects, 
the general solution is:
\begin{equation}
g_{\mu\nu}^{\rm quantum+gravity}=e^{\alpha\Phi}g_{\mu\nu}^{\rm gravity}
\end{equation}
\section{QUANTUM EFFECTS NEAR THAT REGIONS OF THE SPACE--TIME WHERE 
GRAVITY IS LARGE}
In the previous section we have derived the exact solution of our toy field 
equations in terms of the classical solution, i.e. the solution  of Einstein's 
equation. Now, in this section we shall use that solution to examine the 
quantum effects near that regions of the space--time where the gravitational 
effects of matter are large. Black holes and the initial singularity are two 
examples we consider.
\subsection{Quantum Black Hole}
\begin{figure}
\begin{center}
\unitlength 1mm
\linethickness{0.4pt}
\begin{picture}(82.67,84.00)
\put(10.00,80.00){\vector(0,1){0.2}}
\put(10.00,10.00){\line(0,1){70.00}}
\put(80.00,10.00){\vector(1,0){0.2}}
\put(10.33,10.00){\line(1,0){69.67}}
\put(35.00,10.00){\line(0,1){65.00}}
\multiput(10.33,10.00)(0.11,0.35){8}{\line(0,1){0.35}}
\multiput(11.17,12.80)(0.11,0.34){8}{\line(0,1){0.34}}
\multiput(12.07,15.49)(0.12,0.32){8}{\line(0,1){0.32}}
\multiput(13.03,18.05)(0.11,0.27){9}{\line(0,1){0.27}}
\multiput(14.04,20.49)(0.12,0.26){9}{\line(0,1){0.26}}
\multiput(15.11,22.81)(0.11,0.22){10}{\line(0,1){0.22}}
\multiput(16.24,25.01)(0.12,0.21){10}{\line(0,1){0.21}}
\multiput(17.43,27.09)(0.11,0.18){11}{\line(0,1){0.18}}
\multiput(18.67,29.05)(0.12,0.17){11}{\line(0,1){0.17}}
\multiput(19.97,30.89)(0.11,0.14){12}{\line(0,1){0.14}}
\multiput(21.33,32.61)(0.12,0.13){12}{\line(0,1){0.13}}
\multiput(22.75,34.20)(0.11,0.11){13}{\line(0,1){0.11}}
\multiput(24.22,35.68)(0.13,0.11){12}{\line(1,0){0.13}}
\multiput(25.75,37.04)(0.14,0.11){11}{\line(1,0){0.14}}
\multiput(27.34,38.27)(0.18,0.12){15}{\line(1,0){0.18}}
\multiput(30.00,40.00)(0.50,0.10){5}{\line(1,0){0.50}}
\put(32.50,40.50){\line(1,0){5.00}}
\multiput(37.50,40.50)(0.50,-0.10){5}{\line(1,0){0.50}}
\multiput(40.00,40.00)(0.94,0.08){2}{\line(1,0){0.94}}
\multiput(41.88,40.16)(0.62,0.10){3}{\line(1,0){0.62}}
\multiput(43.75,40.47)(0.46,0.11){4}{\line(1,0){0.46}}
\multiput(45.58,40.93)(0.36,0.12){5}{\line(1,0){0.36}}
\multiput(47.40,41.52)(0.26,0.11){7}{\line(1,0){0.26}}
\multiput(49.19,42.26)(0.22,0.11){8}{\line(1,0){0.22}}
\multiput(50.96,43.14)(0.19,0.11){9}{\line(1,0){0.19}}
\multiput(52.70,44.17)(0.17,0.12){10}{\line(1,0){0.17}}
\multiput(54.43,45.34)(0.15,0.12){11}{\line(1,0){0.15}}
\multiput(56.13,46.65)(0.13,0.11){13}{\line(1,0){0.13}}
\multiput(57.80,48.11)(0.12,0.11){14}{\line(1,0){0.12}}
\multiput(59.45,49.71)(0.12,0.12){14}{\line(0,1){0.12}}
\multiput(61.08,51.45)(0.11,0.13){14}{\line(0,1){0.13}}
\multiput(62.69,53.34)(0.11,0.15){14}{\line(0,1){0.15}}
\multiput(64.27,55.37)(0.12,0.17){13}{\line(0,1){0.17}}
\multiput(65.83,57.54)(0.12,0.18){13}{\line(0,1){0.18}}
\multiput(67.37,59.86)(0.12,0.19){13}{\line(0,1){0.19}}
\multiput(68.88,62.32)(0.11,0.20){13}{\line(0,1){0.20}}
\multiput(70.37,64.92)(0.11,0.21){13}{\line(0,1){0.21}}
\multiput(71.84,67.67)(0.11,0.22){13}{\line(0,1){0.22}}
\multiput(73.28,70.56)(0.11,0.25){18}{\line(0,1){0.25}}
\put(115.00,11.00){\makebox(0,0)[cc]{radial distance from the black hole}}
\put(10.00,84.00){\makebox(0,0)[cc]{conformal factor}}
\put(35.00,7.67){\makebox(0,0)[cc]{$r_s$}}
\end{picture}
\caption{Conformal factor for a black hole}
\label{fig1}
\end{center}
\end{figure}
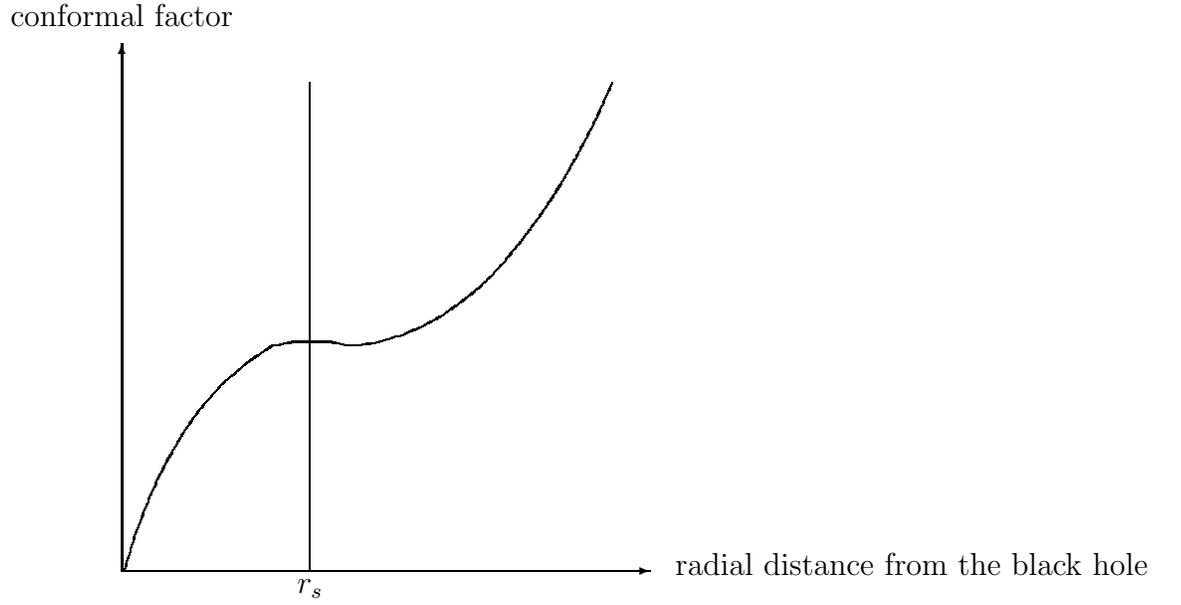

For a spherically symmetric black hole we have
\begin{equation}
g_{\mu\nu}^{\rm gravity}=\left (
\begin{array}{cccc}
1-r_s/r&0&0&0\\
0&\frac{-1}{1-r_s/r}&0&0\\
0&0&-r^2&0\\
0&0&0&-r^2\sin^2\theta
\end{array}
\right )
\end{equation}
Using the fact that the Ricci scalar is zero for the above metric and the 
transformation properties
of the Ricci scalar under conformal transformations we have:
\begin{equation}
\Phi=\frac{\Box\sqrt{|{\cal R}|}}{\sqrt{|{\cal R}|}}
\end{equation}
\begin{equation}
{\cal R}=3\alpha e^{-\alpha\Phi}\left [ \stackrel{-}{\Box}\Phi+\frac{\alpha}{2} 
\overline{\nabla}_\mu \Phi\overline{\nabla}^\mu \Phi\right ]
\end{equation}
The above equations are in fact a differential equation for the conformal factor
and  can be solved for different regimes, giving the following solution:
\begin{equation}
g_{\mu\nu}^{\rm quantum+gravity}=g_{\mu\nu}^{\rm gravity}\times\left \{
\begin{array}{ll}
\exp(-\alpha r_s/r^3)&r\rightarrow 0\\
{\rm Constant}&r\rightarrow r_s\\
\exp(r^2/3\alpha)&r\rightarrow\infty
\end{array}\right .
\end{equation}
The conformal factor is plotted in figure (\ref{fig1}). It can be seen 
that the above conformal factor does not remove the metric singularity at $r=0$.
\subsection{Initial Singularity}
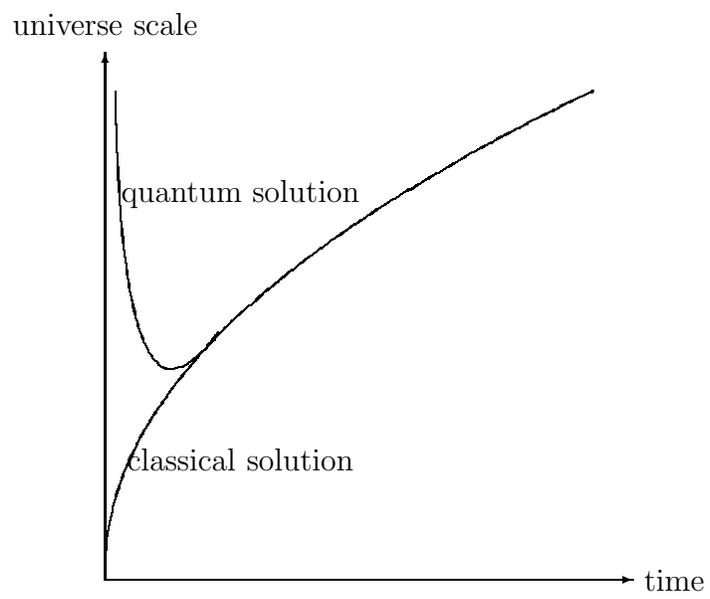
\begin{figure}
\begin{center}
\unitlength 1mm
\linethickness{0.4pt}
\begin{picture}(83.67,84.00)
\put(10.00,80.00){\vector(0,1){0.2}}
\put(10.00,10.00){\line(0,1){70.00}}
\put(80.00,10.00){\vector(1,0){0.2}}
\put(10.00,10.00){\line(1,0){70.00}}
\multiput(75.00,75.00)(-0.24,-0.11){19}{\line(-1,0){0.24}}
\multiput(70.51,72.83)(-0.23,-0.11){19}{\line(-1,0){0.23}}
\multiput(66.19,70.65)(-0.22,-0.11){19}{\line(-1,0){0.22}}
\multiput(62.02,68.47)(-0.21,-0.11){19}{\line(-1,0){0.21}}
\multiput(58.01,66.29)(-0.20,-0.12){19}{\line(-1,0){0.20}}
\multiput(54.17,64.10)(-0.19,-0.12){19}{\line(-1,0){0.19}}
\multiput(50.48,61.91)(-0.19,-0.12){19}{\line(-1,0){0.19}}
\multiput(46.95,59.72)(-0.18,-0.12){19}{\line(-1,0){0.18}}
\multiput(43.59,57.52)(-0.17,-0.12){19}{\line(-1,0){0.17}}
\multiput(40.39,55.31)(-0.16,-0.12){19}{\line(-1,0){0.16}}
\multiput(37.34,53.11)(-0.15,-0.12){19}{\line(-1,0){0.15}}
\multiput(34.46,50.90)(-0.14,-0.12){19}{\line(-1,0){0.14}}
\multiput(31.73,48.68)(-0.13,-0.12){19}{\line(-1,0){0.13}}
\multiput(29.17,46.46)(-0.13,-0.12){19}{\line(-1,0){0.13}}
\multiput(26.77,44.24)(-0.12,-0.12){19}{\line(-1,0){0.12}}
\multiput(24.53,42.01)(-0.12,-0.12){18}{\line(0,-1){0.12}}
\multiput(22.45,39.78)(-0.11,-0.13){17}{\line(0,-1){0.13}}
\multiput(20.52,37.55)(-0.12,-0.15){15}{\line(0,-1){0.15}}
\multiput(18.76,35.31)(-0.11,-0.16){14}{\line(0,-1){0.16}}
\multiput(17.16,33.07)(-0.11,-0.17){13}{\line(0,-1){0.17}}
\multiput(15.72,30.82)(-0.12,-0.20){11}{\line(0,-1){0.20}}
\multiput(14.44,28.57)(-0.11,-0.23){10}{\line(0,-1){0.23}}
\multiput(13.32,26.32)(-0.12,-0.28){8}{\line(0,-1){0.28}}
\multiput(12.37,24.06)(-0.11,-0.32){7}{\line(0,-1){0.32}}
\multiput(11.57,21.80)(-0.11,-0.38){6}{\line(0,-1){0.38}}
\multiput(10.93,19.53)(-0.12,-0.57){4}{\line(0,-1){0.57}}
\multiput(10.45,17.26)(-0.11,-0.76){3}{\line(0,-1){0.76}}
\multiput(10.13,14.99)(-0.08,-1.14){2}{\line(0,-1){1.14}}
\put(9.98,12.71){\line(0,-1){2.71}}
\multiput(25.00,43.00)(-0.12,-0.16){9}{\line(0,-1){0.16}}
\multiput(23.94,41.57)(-0.11,-0.13){9}{\line(0,-1){0.13}}
\multiput(22.93,40.38)(-0.12,-0.12){8}{\line(-1,0){0.12}}
\multiput(21.95,39.43)(-0.16,-0.12){6}{\line(-1,0){0.16}}
\multiput(21.02,38.71)(-0.22,-0.12){4}{\line(-1,0){0.22}}
\multiput(20.13,38.24)(-0.42,-0.12){2}{\line(-1,0){0.42}}
\put(19.29,38.00){\line(-1,0){0.80}}
\multiput(18.48,38.01)(-0.25,0.08){3}{\line(-1,0){0.25}}
\multiput(17.72,38.25)(-0.14,0.10){5}{\line(-1,0){0.14}}
\multiput(17.01,38.73)(-0.11,0.12){6}{\line(0,1){0.12}}
\multiput(16.33,39.45)(-0.11,0.16){6}{\line(0,1){0.16}}
\multiput(15.70,40.41)(-0.12,0.24){5}{\line(0,1){0.24}}
\multiput(15.11,41.61)(-0.11,0.29){5}{\line(0,1){0.29}}
\multiput(14.56,43.05)(-0.10,0.34){5}{\line(0,1){0.34}}
\multiput(14.05,44.73)(-0.12,0.48){4}{\line(0,1){0.48}}
\multiput(13.59,46.64)(-0.11,0.54){4}{\line(0,1){0.54}}
\multiput(13.17,48.80)(-0.09,0.60){4}{\line(0,1){0.60}}
\multiput(12.79,51.19)(-0.11,0.88){3}{\line(0,1){0.88}}
\multiput(12.46,53.83)(-0.10,0.96){3}{\line(0,1){0.96}}
\multiput(12.16,56.70)(-0.08,1.04){3}{\line(0,1){1.04}}
\multiput(11.91,59.81)(-0.10,1.68){2}{\line(0,1){1.68}}
\multiput(11.71,63.16)(-0.08,1.79){2}{\line(0,1){1.79}}
\multiput(11.54,66.75)(-0.06,1.91){2}{\line(0,1){1.91}}
\put(11.42,70.58){\line(0,1){4.42}}
\put(10.00,84.00){\makebox(0,0)[cc]{universe scale}}
\put(85.67,10.00){\makebox(0,0)[cc]{time}}
\put(28.00,61.33){\makebox(0,0)[cc]{quantum solution}}
\put(28.00,26.00){\makebox(0,0)[cc]{classical solution}}
\end{picture}
\caption{The universe scale v.s. time}
\label{fig2}
\end{center}
\end{figure}

For an isotropic and homogeneous universe we have
\begin{equation}
g_{\mu\nu}^{\rm gravity}=
\left (
\begin{array}{cccc}
1&0&0&0\\
0&-\frac{a^2}{1-kr^2}&0&0\\
0&0&-a^2r^2&0\\
0&0&0&-a^2r^2\sin^2\theta
\end{array}
\right )
\end{equation}
As in the previous subsection the equations governing the conformal factor are:
\begin{equation}
\Phi=\frac{\Box\sqrt{|{\cal R}|}}{\sqrt{|{\cal R}|}}
\end{equation}
\begin{equation}
{\cal R}=e^{-\alpha\Phi}\left [\overline{{\cal R}}+ 3 \alpha\left ( 
\stackrel{-}{\Box}\Phi+\frac{\alpha}{2} \overline{\nabla}_\mu \Phi
\overline{\nabla}^\mu \Phi\right )\right ]
\end{equation}
As $t\rightarrow 0$ one can solve the above equations approaximately:
\begin{equation}
g_{\mu\nu}^{\rm quantum+gravity}=g_{\mu\nu}^{\rm gravity}\exp(\alpha/2t^2)
\end{equation}
so the universe scale is given by:
\begin{equation}
a(t)=a^{\rm classic}(t)\exp(\alpha/4t^2)=\sqrt{t}\exp(\alpha/4t^2)
\end{equation}
As it can be seen easily, the curvature singularity at $t=0$ is removed 
because as time goes to zero, the universe scale goes to infinity. 
The behaviour of the universe scale is ploted in figure (\ref{fig2}).
\section{CONCLUDING REMARKS}
Bohmian quantum theory is a theory based on the idea of bringing back 
the ideas of causality and the dogma of formulation of physics in 
terms of motion of particles, to quantum theory. It is a satisfactory 
and successful theory. It is not only a causal theory in terms of particles
trajectories, but also presents a beautiful explanation of how wavefunction
collapse happens and what is the meaning of the uncertainty relations. The
key stone of Bohm's theory is the {\it quantum potential\/}. Any particle is acted 
upon by a quantum force derived from the quantum potential. Quantum potential is 
itself resulted from some self--field of the particle, the wavefunction.
Since quantum potential is related only to the norm of the wavefunction and 
because of the Born's postulate asserting that the ensemble density of the 
particle under consideration is given by the square of the norm of the wavefunction,
quantum potential is resulted from the ensemble density. The nonunderstandable
point of Bohm's theory is just this. How a particle knows about its hypothetical
ensemble? When the hypothetical ensemble is a real one, i.e. when there are
a large number of similar particles just like the particle under
consideration, quantum potential can be understood. It is a kind of interaction
between the particles in the real ensemble. But when one deals with only one particle
the quantum potential is interaction with other hypothetical particles!!!

On the other hand, quantum potential is highly related to the conformal degree
of freedom of the space--time metric. In fact, the presence of the quantum 
force is just like to have a curved space--time which is conformally flat and
the conformal factor is expressed in terms of the quantum potential. In this way
one sees that quantum effects are in fact geometric effects. Geometrization of 
quantum theory can be done successfully. But still there is the problem of ensemble
noted above.

In this paper we have shown that if one tries to geometrize the quantum effects
in a purely metric way, the ensemble problem would be overcommed. In addition
it provides the framework for bringing in the purely quantum gravity effects.

We presented a toy model, and investigate its solutions. It is shown that the initial 
singularity is removed by quantum effects.

\end{document}